\documentclass
[%
 reprint,
 amsmath,amssymb,
 aps,
]{revtex4-2}

\usepackage{graphicx}
\usepackage{dcolumn}
\usepackage{bm}
\usepackage{hyperref}
\usepackage{csquotes}
\usepackage{xcolor}
\usepackage{ulem}
\usepackage{pgfplots}
\usepackage{float}

\begin{document}

\title{Holographic dark energy in Einstein-Cartan spin cosmology}

\author{Yongjun Yun}
\affiliation{Graduate School Department of Physics, Daejin University, Pocheon 11159, Korea}

\author{Jungjai Lee}
\email{jjlee@daejin.ac.kr}
\affiliation{Graduate School Department of Physics, Daejin University, Pocheon 11159, Korea \\
School of Physics, Korea Institute for Advanced Study, Seoul 02455, Korea}

\begin{abstract}
We show that, in Einstein-Cartan theory, the torsion scalar modifies the dust-like equation of state of non-interacting holographic dark energy with the Hubble radius as the infrared cutoff, enabling late-time cosmic acceleration within the low-redshift domain of validity, which is not realized in the corresponding model in general relativity. Using the same holographic dark energy background, we further derive the torsion-induced modification of the cosmic distance duality relation, $d_L=d_A(1+z)^2(1+\eta)$.
In the late-time universe, the magnitude of the deviation parameter, $|\eta|$, is small but non-vanishing, reflecting the cumulative effect of torsion along the light-propagation path.
\end{abstract}
\maketitle

\section{Introduction}
Observations of Type Ia supernovae~\cite{1,2}, the cosmic microwave background~\cite{3,4}, baryon acoustic oscillations~\cite{5,6}, and large-scale structure~\cite{7,8} indicate that the universe is currently undergoing accelerated expansion.
Although the $\Lambda$CDM model provides the simplest explanation, the cosmological constant does not resolve the fine-tuning problem of vacuum energy~\cite{9,10} or the cosmic coincidence problem~\cite{11}.
In addition, the Hubble tension~\cite{12,13,14}, the $S_{8}$ tension~\cite{15}, and recent DESI results~\cite{16} continue to motivate alternatives to the standard cosmological framework.
In particular, models with an interaction between dark matter and dark energy face difficulties in alleviating both tensions simultaneously~\cite{17,18}.
These considerations motivate the exploration of alternative dark energy models in which late-time cosmic acceleration can arise without an explicit interaction between dark matter and dark energy.

Holographic dark energy provides one such possibility.
Cohen et al.~\cite{19} suggested that vacuum energy in a region of finite size $L$ should not exceed the mass of a black hole of the same size, leading to a relation between an ultraviolet (UV) cutoff and an infrared (IR) cutoff.
This UV/IR mixing forms the basis of the holographic principle.
The underlying origin of holographic dark energy remains an open question. In the present work, we adopt it as a phenomenological framework motivated by this UV/IR relation.

In general relativity, non-interacting holographic dark energy with the future event horizon as the IR cutoff can drive cosmic acceleration~\cite{20}, but this choice suffers from causality and circularity problems~\cite{21}.
By contrast, when the Hubble radius is adopted as the IR cutoff, the non-interacting model behaves like dust and cannot drive cosmic acceleration~\cite{22,23}.
An interaction between dark matter and dark energy can modify this dust-like behavior, but the corresponding interacting construction does not admit a non-interacting limit~\cite{24}.

\begin{figure}[htbp]
\centering
\includegraphics[width=\linewidth]{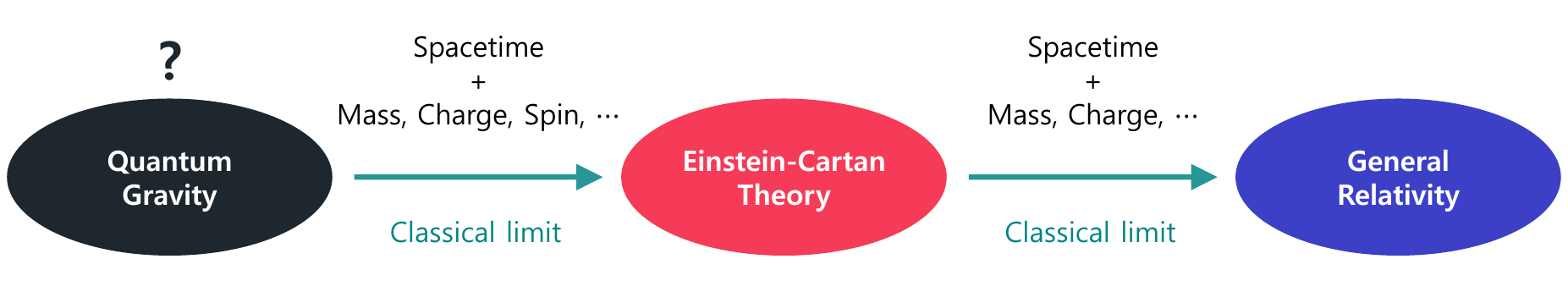} 
\caption{Although a complete theory of quantum gravity has not yet been established, we consider the possibility that its classical limit may admit an Einstein-Cartan description before general relativity is recovered in the torsion-free limit.\label{intro.}}
\end{figure}

These limitations motivate us to ask whether the dust-like behavior can instead be modified by the spacetime geometry itself, without introducing an interaction between dark matter and dark energy.
Intrinsic spin is a fundamental quantum property of matter, and Einstein-Cartan theory provides a natural framework in which spin is incorporated into spacetime geometry through torsion~\cite{25,26,27}.
A possible intermediate classical role of Einstein-Cartan theory between quantum gravity and general relativity is schematically illustrated in Fig.~\ref{intro.}. 
We therefore investigate whether torsion can modify the dust-like behavior of non-interacting holographic dark energy with the Hubble radius as the IR cutoff and thereby allow late-time cosmic acceleration.
The present Einstein-Cartan spin cosmology framework is intended as an effective description of the late-time universe rather than as a unified model of the entire cosmic history. Accordingly, comparisons with observables probing substantially higher redshifts require particular caution.

In this Letter, we derive the Friedmann-like equations for a semiclassical Weyssenhoff spin fluid in the presence of a torsion scalar.
We show that the torsion scalar shifts the equation of state of holographic dark energy from its dust-like value toward negative values, allowing late-time cosmic acceleration while the dark sector remains non-interacting.
Using the same holographic dark energy background, we further derive the torsion-induced modification of the cosmic distance duality relation and examine its possible observational imprint on cosmological distance measurements.

The remainder of this Letter is organized as follows.
In Section~II, we briefly introduce Einstein-Cartan theory.
In Section~III, we apply this theory to an FLRW background.
In Section~IV, we investigate late-time cosmic acceleration driven by holographic dark energy.
In Section~V, we analyze the cosmic distance duality relation in the resulting background.
Finally, we summarize our results in Section~VI.
Throughout this work, we use natural units $c=\hbar=1$.

\section{Einstein-Cartan Gravity}
In Einstein-Cartan theory, the total action is given by
\begin{equation} \label{action}
    S = \frac{1}{2\kappa} \int d^4x \sqrt{-g} R + \int d^4x \sqrt{-g} \mathcal{L}_{m},
\end{equation}
where $R$ is the Ricci-Cartan scalar, $\mathcal{L}_{m}$ is the matter Lagrangian, $g = \det (g_{\mu\nu})$, and $\kappa = 8 \pi G$.
Imposing the metric-compatibility condition, the affine connection can be decomposed as $\Gamma^{\rho}_{\mu\nu} = \tilde{\Gamma}^{\rho}_{\mu\nu} - K_{\mu\nu}{}^{\rho}$, where $\tilde{\Gamma}^{\rho}_{\mu\nu}$ is the Levi-Civita connection and $K_{\mu\nu}{}^{\rho}$ is the contorsion tensor.
The latter is related to the torsion tensor $S_{\mu\nu}{}^{\rho} = \Gamma^{\rho}_{[\mu\nu]}$ by $K_{\mu\nu}{}^{\rho} = - S_{\mu\nu}{}^{\rho} - S^{\rho}{}_{\mu\nu} - S^{\rho}{}_{\nu\mu}$.

Varying~\eqref{action} with respect to the contorsion tensor gives the Cartan field equations
\begin{equation} \label{Cartan eq}
    S_{\lambda\nu}{}^{\mu} + \delta_{\lambda}^{\mu} S_{\nu\alpha}{}^{\alpha} - \delta_{\nu}^{\mu} S_{\lambda\alpha}{}^{\alpha} = \kappa s_{\lambda\nu}{}^{\mu},
\end{equation}
where the spin tensor $s_{\lambda\nu}{}^{\mu}$ is defined as
\begin{equation} \label{spin tensor}
    s_{\lambda\nu}{}^{\mu} = \frac{1}{\sqrt{-g}} \frac{\delta \left( \sqrt{-g} \mathcal{L}_{m} \right)}{\delta K_{\mu}{}^{\nu\lambda}}.
\end{equation}
The Cartan field equations algebraically relate torsion to the spin tensor of matter, implying that torsion does not constitute an independent propagating degree of freedom.
Varying~\eqref{action} with respect to the metric tensor gives the Einstein-Cartan field equations
\begin{equation} \label{EC eq}
    R_{\mu\nu} - \frac{1}{2} g_{\mu\nu} R = \kappa T_{\mu\nu},
\end{equation}
where $R_{\mu\nu}$ is the Ricci-Cartan tensor and $T_{\mu\nu}$ is the canonical energy-momentum tensor.
Using~\eqref{Cartan eq}, the latter can be expressed as
\begin{equation} \label{canonical e-m}
    T_{\mu\nu} = \tilde{T}_{\mu\nu} - \left( \nabla_{\lambda} + 2 S_{\lambda\sigma}{}^{\sigma} \right) \left( -s_{\mu\nu}{}^{\lambda} + s_{\nu}{}^{\lambda}{}_{\mu} - s^{\lambda}{}_{\mu\nu} \right),
\end{equation}
where $\tilde{T}_{\mu\nu}$ is the energy-momentum tensor
\begin{equation} \label{e-m}
    \tilde{T}_{\mu\nu} = -\frac{2}{\sqrt{-g}} \frac{\delta \left( \sqrt{-g} \mathcal{L}_{m} \right)}{\delta g^{\mu\nu}}.
\end{equation}
The second term on the right-hand side of~\eqref{canonical e-m} represents the correction due to the spin. 
In the torsion-free limit,~\eqref{EC eq} reduces to the Einstein field equations.

\section{Einstein-Cartan Spin Cosmology} 
The Cartan field equations~\eqref{Cartan eq} require a specification of the spin source that algebraically determines torsion.
Rather than modeling the microscopic dynamics of individual fermions, we adopt a semiclassical Weyssenhoff spin fluid to describe their intrinsic spin at macroscopic scales.
The spin tensor takes the form~\cite{28}
\begin{equation} \label{spin density}
    s_{\mu\nu}{}^{\alpha} = s_{\mu\nu} u^{\alpha},
\end{equation}
where $s_{\mu\nu}$ is the spin density of the fluid and $u^{\alpha}$ is its four-velocity.
Under the Frenkel condition $s_{\mu\nu} u^\nu = 0$~\cite{29}, the canonical energy-momentum tensor~\eqref{canonical e-m} is given by~\cite{30}
\begin{equation} \label{canonical 2}
    T_{\mu\nu} = \left( \rho + p \right) u_{\mu} u_{\nu} + p g_{\mu\nu} - 2 u^{\lambda} u_{\nu} \tilde{\nabla}_{\alpha} \left( s_{\mu\lambda} u^{\alpha} \right),
\end{equation}
where $\rho$ is the energy density of the fluid, $p$ is its pressure, and $\tilde{\nabla}_{\alpha}$ is the Levi-Civita covariant derivative.

For randomly oriented fermion spins in the fluid, the linear spin average vanishes, whereas the quadratic spin contribution remains non-zero~\cite{26,27,30}, namely
\begin{equation} \label{square spin}
    \langle s_{\mu\nu} \rangle = 0, \qquad \langle s_{\mu\nu} s^{\mu\nu} \rangle = \frac{1}{2} s^{2},
\end{equation}
where $s^2$ is the square of the spin density of the fluid.
Using~\eqref{EC eq},~\eqref{spin density}, and~\eqref{canonical 2}, we obtain the following field equations
\begin{equation} \label{EC eq 2}
	\tilde{G}_{\mu\nu} = \kappa \left( \rho + p - \frac{1}{2} \kappa s^{2} \right) u_{\mu} u_{\nu} + \kappa \left( p - \frac{1}{4} \kappa s^{2} \right) g_{\mu\nu},
\end{equation}
where $\tilde{G}_{\mu\nu}$ is the Einstein tensor.

We consider a spatially homogeneous and isotropic universe.
Although the torsion tensor has 24 independent components, the cosmological principle restricts the non-vanishing components to $S_{[123]}$ and $S_{01}{}^1=S_{02}{}^2=S_{03}{}^3$, which depend only on cosmic time $t$~\cite{31}.
In this work, we restrict our analysis to the trace sector, which can be parametrized by a time-dependent function $\Phi(t)$ as (see~\cite{32} for details)
\begin{equation} \label{torsion scalar}
    S_{\mu\nu\rho} = \Phi(t) h_{\rho[\mu} u_{\nu]},
\end{equation}
where $h_{\mu\nu}$ is the projection tensor orthogonal to $u^{\mu}$.
In the comoving frame, where $u^{\mu} = ( 1,0,0,0 )$, this gives $S_{01}{}^{1} = S_{02}{}^{2} = S_{03}{}^{3} = \Phi / 2$.
From~\eqref{Cartan eq},~\eqref{spin density},~\eqref{square spin}, and~\eqref{torsion scalar}, we find the relation
\begin{equation} \label{spin-torsion}
    s^{2} = 12 M_{p}^{4} \Phi^{2},
\end{equation}
where $M_{p} = 1 / \sqrt{8 \pi G}$ is the reduced Planck mass.
Thus, the torsion scalar $\Phi$ is not an independent propagating degree of freedom, nor is it introduced as the time derivative of a scalar field.
Rather, its magnitude is determined algebraically by the spin density of the fluid.

In this context, we can employ the flat FLRW metric
\begin{equation} \label{FLRW}
	ds^{2} = -dt^{2} + a^{2}(t) \left( dr^{2} + r^{2} d\theta^{2} + r^{2} \sin^{2}\theta d\varphi^{2} \right),
\end{equation}
where $a(t)$ is the scale factor.
In this metric, the Einstein-Cartan field equations~\eqref{EC eq 2} lead to the Friedmann-like equations
\begin{equation} \label{Friedmann-like eq 1}
	\left( \frac{\dot{a}}{a} \right)^{2} = \frac{1}{3 M_{p}^{2}} \left( \rho - 3 M_{p}^{2} \Phi^{2} \right)
\end{equation}
and
\begin{equation} \label{Friedmann-like eq 2}
	\frac{\ddot{a}}{a} = -\frac{1}{6 M_{p}^{2}} \left( \rho + 3 p - 12 M_{p}^{2} \Phi^{2} \right).
\end{equation}
In the torsion-free limit,~\eqref{Friedmann-like eq 1} and~\eqref{Friedmann-like eq 2} reduce to the standard Friedmann equations.

Since our primary interest is late-time cosmic acceleration, we neglect radiation and consider two energy components, matter and dark energy, denoted by the subscripts $m$ and $X$, respectively.
The total energy density and pressure are given by $\rho = \rho_{m} + \rho_{X}$ and $p = p_{m} + p_{X}$.
We define the Hubble parameter $H = \dot{a} / a$, the critical density $\rho_{c} = 3 M_{p}^{2} H^{2}$, and the density parameters
$\Omega_{m} = \rho_{m} / \rho_{c}$ and $\Omega_{X} = \rho_{X} / \rho_{c}$.
In terms of the dimensionless quantities, the first Friedmann-like equation~\eqref{Friedmann-like eq 1} can be rewritten as
\begin{equation} \label{Friedmann-like eq 3}
    \Omega_{m} + \Omega_{X} - \left( \frac{\Phi}{H} \right)^{2} = 1.
\end{equation}
We further define the deceleration parameter $q = - \ddot{a} a / \dot{a}^{2}$ and the equations of state $\omega_{m} = p_{m} / \rho_{m}$ and $\omega_{X} = p_{X} / \rho_{X}$.
The second Friedmann-like equation~\eqref{Friedmann-like eq 2} can then be expressed as
\begin{equation} \label{deceleration parameter}
	q = \frac{1}{2} \Omega_{m} \left( 1 + 3 \omega_{m} \right) + \frac{1}{2} \Omega_{X} \left( 1 + 3 \omega_{X} \right) - 2 \left( \frac{\Phi}{H} \right)^{2}.
\end{equation}

Assuming adiabatic expansion of the universe and conservation of the total particle number, we have the relation $dn / n = d\rho / ( \rho + p )$, where $n$ is the particle number density.
For an unpolarized fermion fluid, the spin density and particle number density satisfy
$s^{2} = n^{2} / 8$~\cite{28,30}, which, together with~\eqref{spin-torsion}, gives $n^{2}= 96 M_{p}^{4} \Phi^{2}$.
Combining the Friedmann-like equations~\eqref{Friedmann-like eq 1} and~\eqref{Friedmann-like eq 2} with these relations, we obtain the continuity equation
\begin{equation} \label{continuity eq}
    \dot{\rho} + 3 H \left( \rho + p \right) = 0
\end{equation}
and the evolution equation
\begin{equation} \label{Phi eq}
    \dot{\Phi} + 3 H \Phi = 0.
\end{equation}
The latter governs the background evolution of the torsion scalar but does not constitute an independent equation of motion.

In the absence of a phenomenological interaction between dark matter and dark energy, the continuity equation~\eqref{continuity eq} separates into
\begin{equation} \label{continuity eq m}
	\dot{\rho}_{m} + 3 H \rho_{m} \left( 1 + \omega_{m} \right) 
    = 0
\end{equation}
and
\begin{equation} \label{continuity eq X}
	\dot{\rho}_{X} + 3 H \rho_{X} \left( 1 + \omega_{X} \right) 
    = 0.
\end{equation}
For dust with $\omega_{m} = 0$,~\eqref{continuity eq m} gives $\rho_{m} \propto a^{-3}$, while~\eqref{Phi eq} gives
$\Phi \propto a^{-3}$.
However, the torsion contribution to the first Friedmann-like equation~\eqref{Friedmann-like eq 1} is quadratic in $\Phi$, so that the background scalings are
\begin{equation} \label{scaling relations}
    \rho_{m} \propto a^{-3}, \qquad \Phi^{2} \propto a^{-6}.
\end{equation}
Consequently, the torsion contribution is distinct from the dust component already at the background level.
Further distinctions may arise at the perturbative level, but a perturbative analysis is beyond the scope of the present work, which is restricted to the homogeneous and isotropic background.

\section{Holographic Dark Energy and Late-Time Cosmic Acceleration}
Cohen et al.~\cite{19} suggested that vacuum energy in a region of finite size $L$ should not exceed the mass of a black hole of the same size, leading to the upper bound $L^{3} \rho_{X} \lesssim L M_{p}^{2}$.
Saturating this bound gives the holographic dark energy density $\rho_{X} = 3 d^{2} M_{p}^{2} L^{-2}$, where $d$ is a free parameter~\cite{20}.

We examine whether non-interacting holographic dark energy with the Hubble radius as the IR cutoff, $L = H^{-1}$, can drive late-time cosmic acceleration in Einstein-Cartan spin cosmology.
The dark energy density then becomes
\begin{equation} \label{HDE}
    \rho_{X} = 3 d^{2} M_{p}^{2} H^{2}.
\end{equation}
The corresponding density parameter is $\Omega_{X} = d^{2}$.

To assess late-time cosmic acceleration, we determine the equation of state of the dark energy.
The dark energy continuity equation~\eqref{continuity eq X} gives
\begin{equation} \label{EoS}
   \omega_{X} = - 1 - \frac{1}{3} \frac{d \ln{\rho_{X}}}{d \ln{a}}.
\end{equation}
From the first Friedmann-like equation~\eqref{Friedmann-like eq 3} and the holographic constraint~\eqref{HDE}, the dark energy density can be written as
\begin{equation} \label{dark energy density}
    \rho_{X} = \Omega_{X} \rho_{c} = \Omega_{X} \left( \frac{\rho_{m}}{\Omega_{m}} \right) = \rho_{m} \frac{d^{2}}{1 - d^{2} + \left( \frac{\Phi}{H} \right)^{2}}.
\end{equation}
Substituting~\eqref{dark energy density} into~\eqref{EoS} and using~\eqref{deceleration parameter},~\eqref{Phi eq}, and~\eqref{continuity eq m}, we obtain the equation of state
\begin{equation} \label{EoS 2}
    \omega_{X} = \left[ 1 + \frac{1}{1 - d^{2}} \left( \frac{\Phi}{H} \right)^{2} \right] \omega_{m} - \frac{1}{1 - d^{2}} \left( \frac{\Phi}{H} \right)^{2}.
\end{equation}
In the torsion-free limit,~\eqref{EoS 2} reduces to $\omega_{X} = \omega_{m}$, so that for dust with $\omega_{m} = 0$, the dark energy behaves like dust and cannot drive cosmic acceleration.
In general relativity, this dust-like behavior can be modified by introducing an interaction between dark matter and dark energy, but the corresponding interacting model does not admit a non-interacting limit~\cite{24}.
By contrast, in the present Einstein-Cartan framework, the torsion scalar shifts $\omega_{X}$ toward negative values, allowing late-time cosmic acceleration without such an interaction.
In this sense, the torsion scalar can play a role analogous to that of a dark-sector interaction.

As shown in Appendix~A, the standard relation between the scale factor and redshift, $a = ( 1 + z )^{-1}$, remains unchanged despite the presence of the torsion scalar.
We therefore characterize the late-time background evolution in terms of the dimensionless Hubble parameter $E(z) = H(z) / H_{0}$.
Under the holographic constraint~\eqref{HDE}, the first Friedmann-like equation~\eqref{Friedmann-like eq 3} yields the present background constraint
\begin{equation} \label{present torsion}
    \Omega_{m}^{0} + d^{2} - 1 = \left( \frac{\Phi_{0}}{H_{0}} \right)^{2} \geq 0.
\end{equation}
Using the scaling relations~\eqref{scaling relations}, we obtain the background evolution equation
\begin{equation}\label{Ez}
    E^{2}(z) = \frac{\Omega_{m}^{0} (1 + z)^{3} - \left( \Omega_{m}^{0} + d^{2} - 1 \right) (1 + z)^{6}}{1 - d^{2}}.
\end{equation}
The second term in the numerator of~\eqref{Ez} represents the negative torsion contribution.
Its magnitude scales as $( 1 + z )^{6}$ and grows more rapidly toward higher redshift than the matter contribution, which scales as $( 1 + z)^{3}$.

For the representative value $\Omega_{m}^{0} = 0.3$, the background constraint~\eqref{present torsion} requires $d \gtrsim 0.837$, and therefore the limit $d\to0$ lies outside the physical parameter region of the present background solution.
Even in this formal limit,~\eqref{HDE} gives $\rho_{X} \to 0$, so that the dark energy density vanishes.
Since the present model does not include a separate cosmological constant $\Lambda$, it should be regarded as an alternative late-time cosmological framework rather than as a continuous deformation of $\Lambda$CDM recovered in the limit $d\to0$.

A physically meaningful background requires $E^{2}(z) > 0$.
Defining $z_{*}$ by $E^{2}(z_{*}) = 0$, we find the limiting redshift
\begin{equation} \label{limiting redshift}
    z_{*} = \left( \frac{\Omega_{m}^{0}}{\Omega_{m}^{0} + d^{2} - 1} \right)^{1/3} - 1.
\end{equation}
The background solution is therefore restricted to $0 \leq z < z_{*}$.
This restriction arises because the negative torsion contribution grows more rapidly with redshift than the matter contribution.
Accordingly, the present framework should be regarded as an effective description of the late-time universe rather than as a model of the complete cosmic expansion history.

From~\eqref{present torsion} and~\eqref{Ez}, the torsion ratio is given by
\begin{equation} \label{Phi over H}
    \frac{\Phi}{H}(z) = \pm \sqrt{\frac{(1 - d^{2}) \left( \Omega_{m}^{0} + d^{2} - 1 \right) (1 + z)^{3}}{\Omega_{m}^{0} - \left( \Omega_{m}^{0} + d^{2} - 1 \right) (1 + z)^{3}}}.
\end{equation}
Although $E^{2}(z) > 0$ for $0 \leq z < z_{*}$, the weak torsion condition $|\Phi / H|<1$ imposes a more restrictive domain.
Defining $z_{\mathrm{wt}}$ by $|\Phi / H|(z_{\mathrm{wt}}) = 1$, we have
\begin{equation} \label{weak torsion redshift}
    z_{\mathrm{wt}} = \left[ \frac{\Omega_{m}^{0}}{\left( \Omega_{m}^{0} + d^{2} - 1 \right) (2 - d^{2})} \right]^{1/3} - 1.
\end{equation}

Using~\eqref{Ez}, the deceleration parameter $q = - 1 - \dot{H} / H^{2}$ can be expressed as
\begin{equation} \label{qz}
    q(z) = \frac{\frac{1}{2} \Omega_{m}^{0} - 2 \left( \Omega_{m}^{0} + d^{2} - 1 \right) (1 + z)^{3}}{\Omega_{m}^{0} - \left( \Omega_{m}^{0} + d^{2} - 1 \right) (1 + z)^{3}}.
\end{equation}
Rather than adopting a value of $q_{0}^{\Lambda \mathrm{CDM}}$ inferred within the $\Lambda$CDM framework, which assumes a torsion-free background in general relativity, we use a low-redshift cosmographic estimate as a less model-dependent phenomenological reference.
Specifically, we adopt the local cosmographic estimate $q_{0}^{\mathrm{local}} = - 1.08 \pm 0.29$ obtained from the low-redshift Pantheon supernova sample over the interval $0.023 \leq z \leq 0.15$~\cite{33}.

Assuming a Gaussian uncertainty and fixing $\Omega_m^0=0.3$, we employ a simple chi-square statistic
\begin{equation} \label{chi square}
    \chi^{2}(d) = \frac{\left[ q_{0}(d) - q_{0}^{\mathrm{local}} \right]^{2}}{\sigma_{q_{0}^{\mathrm{local}}}^{2}},
\end{equation}
where $q_{0}(d)$ denotes the present deceleration parameter obtained from~\eqref{qz} at $z=0$  and $\sigma_{q_{0}^{\mathrm{local}}} = 0.29$ is the $1 \sigma$ uncertainty in this local cosmographic estimate.
Minimizing~\eqref{chi square} gives
\begin{equation} \label{best}
    d_{\mathrm{best}} = 0.924^{+0.007}_{-0.008},
\end{equation}
where the quoted $1 \sigma$ interval corresponds to $\Delta \chi^{2} = \chi^{2}(d) - \chi^{2}_{\mathrm{min}} = 1$ for one fitted parameter.
The best-fit value is below but close to unity, consistent with previous phenomenological considerations favoring values of $d$ near unity~\cite{23} and numerically close to the theoretical estimate motivated by quantum information theory~\cite{34}.

Using the best-fit value~\eqref{best} with $\Omega_m^0=0.3$,~\eqref{limiting redshift} and~\eqref{weak torsion redshift} give $z_*=0.249$ and $z_{\rm wt}=0.194$, respectively.
These values satisfy
\begin{equation}
0.023\leq z\leq0.15 < z_{\rm wt}=0.194 < z_*=0.249,
\end{equation}
so that the entire cosmographic interval lies within both the weak torsion domain and the physical background domain.
Evaluating~\eqref{Phi over H} at $z=0$ yields
\begin{equation}
    \left|\frac{\Phi_0}{H_0}\right|=0.392^{+0.016}_{-0.019}.
\end{equation}

Substituting~\eqref{Phi over H} into~\eqref{EoS 2}, the equation of state of the dark energy can be written as
\begin{equation} \label{wXz}
\omega_X(z) = -\frac{\left(\Omega_m^0+d^2-1\right)(1+z)^3}{\Omega_m^0-\left(\Omega_m^0+d^2-1\right)(1+z)^3},
\end{equation}
where we have used $\omega_m=0$.
Using the best-fit value~\eqref{best} with $\Omega_m^0=0.3$, we obtain
\begin{equation}
    \omega_X^0=-1.05^{+0.19}_{-0.20}.
\end{equation}
The value $\omega_X^{0} = - 1$, corresponding to a cosmological constant, lies within the $1 \sigma$ interval.
These results indicate that non-interacting holographic dark energy with the Hubble radius as the IR cutoff can drive late-time cosmic acceleration in Einstein-Cartan theory without invoking a phenomenological interaction between dark matter and dark energy.
The present analysis should be regarded as a one-observable, one-parameter consistency estimate. A full multi-probe likelihood analysis involving SNe, BAO, CC, and CMB data would require a consistent extension of the model beyond the present low-redshift domain and is left for future work.

\section{Cosmic Distance Duality Relation in the Present Model}
In this section, we examine the cosmic distance duality relation between the luminosity distance and the angular diameter distance using the holographic dark energy background and the corresponding torsion ratio $\Phi/H$ obtained in the previous section. Combining~(30) and (37), we obtain $\left(\Phi/H\right)^{2}=-(1-d^{2})\omega_{X}$, which directly links the torsion ratio to the equation of state of the holographic dark energy. We therefore examine whether the same torsion background that enables late-time cosmic acceleration also leaves an imprint on cosmological distance measurements.

\begin{figure}[htbp]
\centering
\includegraphics[width=0.75\linewidth]{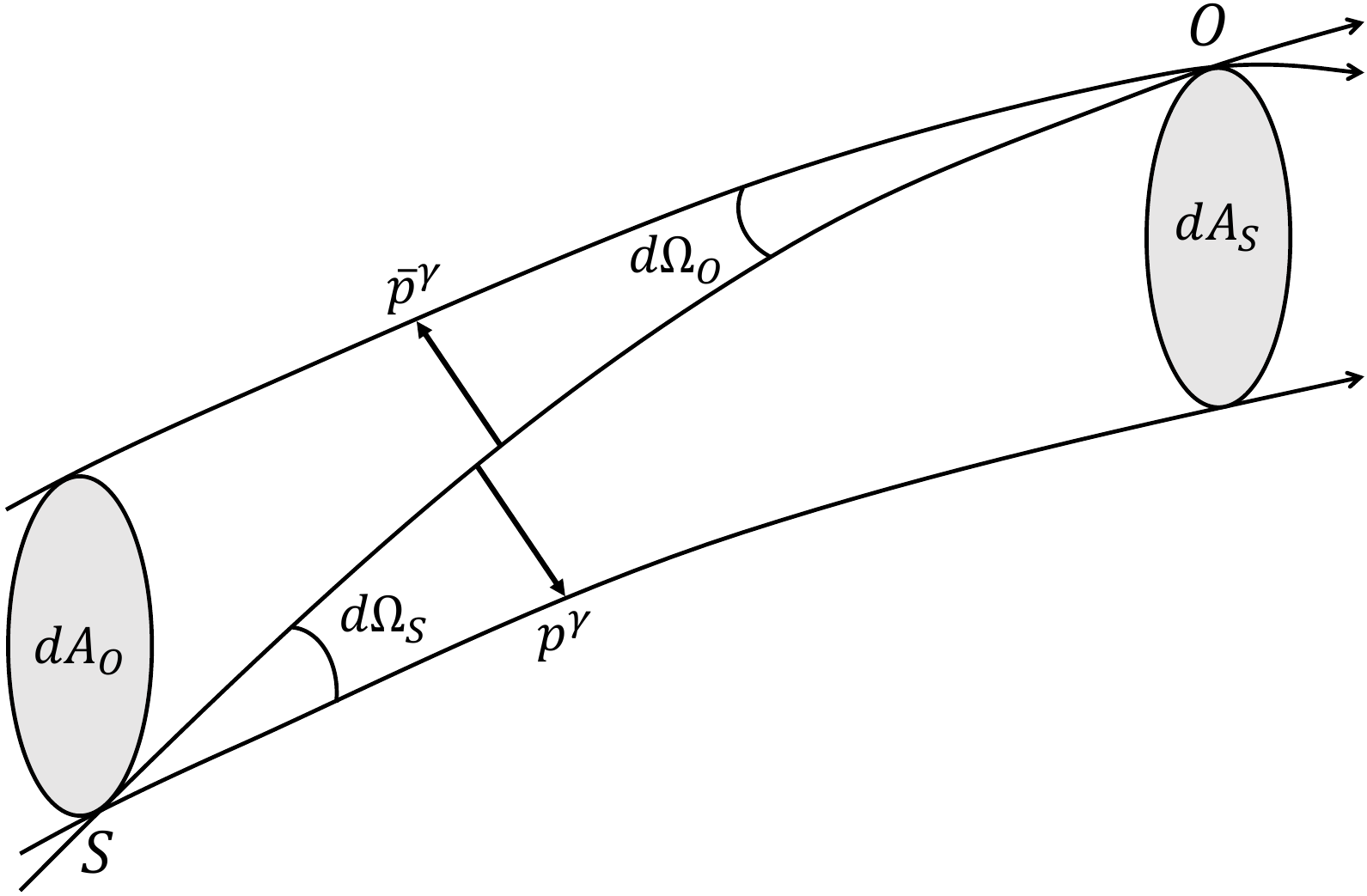} 
\caption{A bundle of null curves diverging from a source $S$ with a solid angle $d\Omega_{S}$ forms a cross-sectional area $dA_{S}$ at an observer $O$.
Conversely, a bundle of null curves converging to $O$ with $d\Omega_{O}$ forms $dA_{O}$ at $S$.
$SO$ is a null curve common to the two bundles.
$p^{\gamma}$ and $\bar{p}^{\gamma}$ are connecting vectors~\cite{35,38}.
\label{null}}
\end{figure}

We consider a system consisting of two null bundles, as shown in Fig.~\ref{null}.
In spacetimes with torsion, the reciprocity theorem between the area distance $r_{S}$ at a source $S$ and the area distance $r_{O}$ at an observer $O$ is modified to (see~\cite{35} for details)
\begin{equation} \label{reciprocity relation 3}
    r_{S}^{2} = r_{O}^{2} \left( 1 + z \right)^{2} (1 + \alpha),
\end{equation}
where $\alpha$ is a function that includes torsion effects. 
In the torsion-free limit,~\eqref{reciprocity relation 3} reduces to the standard reciprocity relation, $r_{S}^{2} = r_{O}^{2} ( 1 + z )^{2}$.

The luminosity distance $d_{L}$ and the angular diameter distance $d_{A}$ are defined as
\begin{equation} \label{cosmological distances}
    d_{L}^{2} = \frac{L}{4 \pi \mathcal{F}_{O}}, \qquad d_{A}^{2} = \frac{dA_{O}}{d\Omega_{O}} = r_{O}^{2},
\end{equation}
where $L$ is the luminosity of the source $S$ and $\mathcal{F}_{O}$ is the flux measured by the observer $O$ with four-velocity $u^{\mu}$.

To obtain the flux, we use the Maxwell Lagrangian
\begin{equation} \label{Maxwell}
    \mathcal{L}_{m} = - \frac{1}{4} g^{\mu\alpha} g^{\nu\beta} F_{\mu\nu} F_{\alpha\beta}.
\end{equation}
In the presence of torsion, minimally coupling the electromagnetic gauge field to the affine connection is incompatible with $U(1)$ gauge invariance~\cite{36}.
As in Ref.~\cite{37}, we retain the gauge-invariant definition $F = dA$, so that $F_{\mu\nu} = \partial_{\mu} A_{\nu} - \partial_{\nu} A_{\mu}$.
Since the Maxwell Lagrangian is independent of the affine connection, the spin tensor~\eqref{spin tensor} vanishes, $s_{\mu\nu}{}^{\rho} = 0$.

Using~\eqref{canonical e-m} and applying the geometric optics approximation~\eqref{geometric optics}, the canonical energy-momentum tensor of radiation takes the form
\begin{equation} \label{canonical energy-momentum tensor radiation}
    T_{\mu\nu}^{\mathrm{rad}} = F_{\mu\rho} F_{\nu}^{\phantom{\nu}\rho} - \frac{1}{4} g_{\mu\nu} F_{\rho\sigma} F^{\rho\sigma} \simeq \mathcal{A}^{2} k_{\mu} k_{\nu},
\end{equation}
where $\mathcal{A}_{\mu}$ is the amplitude of the electromagnetic wave, $\mathcal{A}^{2} = \mathcal{A}_{\mu} \mathcal{A}^{\mu}$, and $k^{\mu} = g^{\mu\nu} k_{\nu}$ is the wave vector.
For the frequency $f = - k_{\alpha} u^{\alpha}$, the flux is given by
\begin{equation} \label{observed flux}
    \mathcal{F} = T_{\mu\nu}^{\mathrm{rad}} u^{\mu} u^{\nu} \simeq \mathcal{A}^{2} f^{2}.
\end{equation}

We treat radiation as a test field decoupled from the matter sector and assume that its canonical energy-momentum tensor is covariantly conserved, $\nabla_{\mu} T_{\mathrm{rad}}^{\mu\nu} = 0$.
This leads to
\begin{equation} \label{divergence null}
    \mathcal{A}^{2} k^{\beta} \nabla_{\beta} k^{\alpha} = -k^{\alpha} k^{\beta} \nabla_{\beta} \mathcal{A}^{2} - \mathcal{A}^{2} k^{\alpha} \nabla_{\beta} k^{\beta}.
\end{equation}
The frequencies emitted at the source at $t_S$ and measured by the observer at $t_{O}$ are denoted by $f_{S}$ and $f_{O}$, respectively, so that $f_{S} / f_{O} = 1 + z$.
Combining~\eqref{divergence null} with~\eqref{null eq} and~\eqref{nabla k}, and using~\eqref{observed flux} and~\eqref{null vector}, we obtain the conserved quantity
\begin{equation} \label{cons}
    \frac{(1 + z)^{2} \mathcal{F} dA}{1 + \beta} \simeq \mathrm{constant},
\end{equation}
where $\beta$ is the torsion-induced correction
\begin{equation} \label{b}
\begin{aligned}
    \beta = \int_{t_{S}}^{t_{O}} dt \Big( & 2 S_{\mu\alpha\nu} n^{\mu} u^{\alpha} u^{\nu} + 2 S_{\mu\alpha\nu} n^{\mu} u^{\alpha} n^{\nu} \\ & + 2 S_{\rho\alpha}{}^{\alpha} k^{\rho} + 4 S_{\mu\nu\rho} k^{\mu} N^{\nu} k^{\rho} \Big).
\end{aligned}
\end{equation}

We define the luminosity
\begin{equation}
    L = \int dA \frac{(1 + z)^{2} \mathcal{F}}{1 + \beta}.
\end{equation}
For isotropic emission, the flux measured on a unit sphere centered at the source is $\mathcal{F}_{S} = L ( 1 + \beta ) / 4 \pi$.
Since $dA_{S} = d\Omega_{S}$ on the unit sphere, the conserved quantity~\eqref{cons} at the source $S$ becomes $\mathcal{F}_{S} d\Omega_{S} / ( 1 + \beta ) = \mathrm{constant}$.
At the observer $O$, it takes the form $( 1 + z )^{2} \mathcal{F}_{O} dA_{S} / ( 1 + \beta ) = \mathrm{constant}$.
Equating the two relations and using the area distance $r_{S}^{2} = dA_{S} / d\Omega_{S}$, the observed flux is given by
\begin{equation} \label{flux by O}
    \mathcal{F}_{O} = \frac{L}{4 \pi r_{S}^{2} \left( 1 + z \right)^{2}} \left( 1 + \beta \right).
\end{equation}

From~\eqref{reciprocity relation 3},~\eqref{cosmological distances}, and~\eqref{flux by O}, we obtain the modified cosmic distance duality relation
\begin{equation} \label{cddl}
    d_{L} = d_{A} \left( 1 + z \right)^{2} \left( 1 + \eta \right),
\end{equation}
where $\eta = \sqrt{( 1 + \alpha ) / ( 1 + \beta )} - 1$ is the deviation parameter~\cite{35}.
Here, $\alpha$ and $\beta$ characterize the torsion effects on the relation between the two area distances and on the observed flux, respectively.
In the torsion-free limit,~\eqref{cddl} reduces to the standard cosmic distance duality relation $d_{L} = d_{A} ( 1 + z )^{2}$.

\begin{figure}[htbp]
\centering
\includegraphics[width=\linewidth]{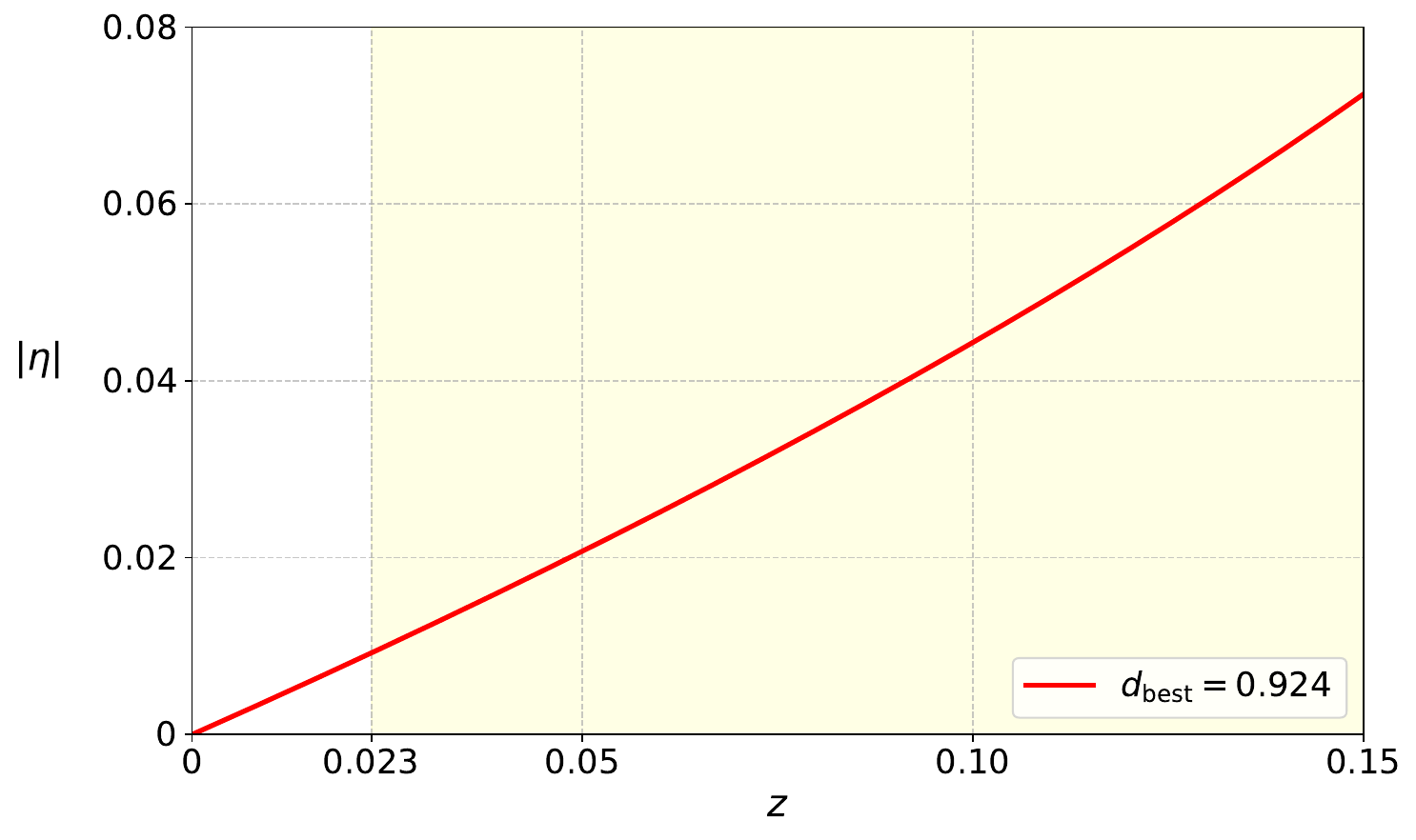}
\caption{Deviation parameter $|\eta|$ as a function of redshift for $d_{\mathrm{best}} = 0.924$ and $\Omega_{m}^{0} = 0.3$.
The shaded region indicates the low-redshift interval $0.023 \leq z \leq 0.15$ used in the local cosmographic analysis.
The curve is shown within the weak torsion domain $|\Phi / H| < 1$, which also lies within the physical background domain $E^{2}(z) > 0$. The increase of $|\eta|$ with redshift reflects the accumulated effect of the torsion scalar along the light path. \label{eta fig}}
\end{figure}

In the weak torsion approximation, $\alpha$ is quadratic in torsion, whereas $\beta$ is linear, so that $\alpha$ can be neglected to leading order.
Using~\eqref{b} together with~\eqref{torsion scalar}, the deviation parameter can be expressed as
\begin{equation} \label{eta 22}
    \eta \approx \frac{1}{2} \left( \alpha - \beta \right) \approx -\frac{1}{2} \beta \approx \int_{t_{S}}^{t_{O}} dt \Phi = \int_{0}^{z} \frac{dz}{1 + z} \frac{\Phi}{H},
\end{equation}
where we have set $z_{S} = z$ and $z_{O} = 0$.
Substituting the background torsion ratio~\eqref{Phi over H} into~\eqref{eta 22}, we arrive at
\begin{equation}
    \eta(z) \approx \pm \frac{2}{3} \sqrt{1 - d^{2}} \left[ \sin^{-1} \left( \xi \sqrt{(1 + z)^{3}} \right) - \sin^{-1} \left( \xi \right) \right],
\end{equation}
where $\xi = \sqrt{( \Omega_{m}^{0} + d^{2} - 1 ) / \Omega_{m}^{0}}$.
Fig.~\ref{eta fig} shows that $|\eta|$ increases monotonically with the source redshift $z$ over the low-redshift interval relevant to the local cosmographic analysis.
This behavior reflects the accumulated effect of the torsion scalar along the light path.
The condition $|\eta|(0) = 0$ does not imply vanishing torsion but follows from the vanishing propagation distance at $z = 0$.
The resulting deviation from the standard cosmic distance duality relation may provide an observational signature of the torsion scalar in cosmological distance measurements.

\section{Conclusions}
Within a low-redshift effective description, we have investigated non-interacting holographic dark energy with the Hubble radius as the infrared cutoff in Einstein-Cartan spin cosmology.
Using a semiclassical Weyssenhoff spin fluid, we derived the Friedmann-like equations in the presence of the torsion scalar $\Phi$.
The Cartan field equations relate $\Phi$ algebraically to the spin density, so that it does not constitute an independent propagating degree of freedom.
Although $\Phi\propto a^{-3}$, its contribution to the Friedmann-like equation is quadratic and scales as $\Phi^{2}\propto a^{-6}$, distinguishing it from dust at the background level.
The torsion scalar shifts the equation of state of the holographic dark energy toward negative values and allows late-time cosmic acceleration while the dark sector remains non-interacting.

Adopting the low-redshift cosmographic estimate of the present deceleration parameter, a simple chi-square analysis gave $d_{\mathrm{best}}=0.924^{+0.007}_{-0.008}$ at $1\sigma$ for the free parameter $d$. 
For this best-fit value, we obtained $\left|\Phi_{0}/H_{0}\right|=0.392^{+0.016}_{-0.019}$ and $\omega_{X}^{0}=-1.05^{+0.19}_{-0.20}$, with the cosmological-constant value $\omega_{X}^{0}=-1$ lying within the $1\sigma$ interval.
The low-redshift interval considered in the local cosmographic analysis lies within both the weak torsion domain and the physical background domain.

Using the same holographic dark energy background, we further showed that the torsion ratio $\Phi/H$, which governs the equation of state of the holographic dark energy, also enters the modified cosmic distance duality relation between the luminosity distance $d_{L}$ and the angular diameter distance $d_{A}$, $d_{L}=d_{A}(1+z)^{2}(1+\eta)$. Within the weak torsion approximation, the magnitude of the deviation parameter, $|\eta|$, increases monotonically with redshift over the low-redshift interval, reflecting the accumulated effect of the torsion scalar along the light path.
This behavior suggests that torsion may leave an observable imprint on cosmological distance measurements.

\acknowledgments
This work was supported by National Research Foundation of Korea (NRF) Grants funded by the Korea Government (MSIT) (No. NRF-2020R1F1A1068410).

\appendix
\section{Light propagation and redshift}
To preserve $U(1)$ gauge invariance in spacetimes with torsion, we retain the gauge-invariant definition $F=dA$, for which the source-free Maxwell equations are $dF=0$ and $d\star F=0$, with the latter equivalent to $\tilde{\nabla}_{\mu}F^{\mu\nu}=0$.
Applying the geometric optics approximation, the electromagnetic field strength takes the form~\cite{38}
\begin{equation} \label{geometric optics}
    F_{\mu\nu} \simeq G'\left(k_{\mu}\mathcal{A}_{\nu}-k_{\nu}\mathcal{A}_{\mu}\right),
\end{equation}
where $G$ is a function of the phase $\phi$, $G' = dG / d\phi$, $\mathcal{A}_{\mu}$ is the wave amplitude, and $k^{\mu}=g^{\mu\nu}\tilde{\nabla}_{\nu}\phi = g^{\mu\nu}\nabla_{\nu}\phi$ is the wave vector.
At leading order, the source-free Maxwell equation yields the eikonal equation
\begin{equation} \label{eikonal eq}
    k_{\mu}k^{\mu}=0.
\end{equation}
Taking the affine covariant derivative of this equation gives $k^{\mu}\nabla_{\nu}k_{\mu}=0$.
Together with $\nabla_{\mu}k_{\nu}-\nabla_{\nu}k_{\mu}=-2S_{\mu\nu}{}^{\rho}k_{\rho}$, this leads to
\begin{equation} \label{null eq}
    k^{\mu}\nabla_{\mu}k_{\nu} = 2S_{\mu\nu\rho} k^{\mu}k^{\rho}.
\end{equation}

We consider a null curve $x^{\mu}(\lambda)$ parameterized by a non-affine parameter $\lambda$ from a source to an observer, with tangent vector $k^{\mu}=dx^{\mu}/d\lambda$.
For the frequency $f=-k_{\alpha}u^{\alpha}$ measured by an observer with four-velocity $u^{\alpha}$, we have the propagation equation
\begin{equation} \label{} \label{propagation equation}
    \frac{Df}{d\lambda} = \frac{dx^\mu}{d\lambda}\nabla_{\mu}f = - k^{\mu}k^{\alpha}\nabla_{\mu}u_{\alpha} - u^{\alpha}k^{\mu}\nabla_{\mu}k_{\alpha},
\end{equation}
where $D/d\lambda$ is the directional affine covariant derivative.
The covariant derivative of $u^\alpha$ can be decomposed as
\begin{equation} \label{1+3 decomposition}
    \nabla_{\mu} u_{\alpha} = \frac{1}{3} \Theta h_{\mu\alpha} + \sigma_{\mu\alpha} + \omega_{\mu\alpha},
\end{equation}
where $\Theta = \nabla_{\gamma} u^{\gamma}$ is the expansion scalar, $\sigma_{\mu\alpha}$ is the symmetric traceless shear tensor, and $\omega_{\mu\alpha}$ is the antisymmetric vorticity tensor.
The expansion scalar satisfies $\Theta = \tilde{\Theta} - 2 S_{\nu\mu}{}^{\mu} u^{\nu}$, with $\tilde{\Theta} = \tilde{\nabla}_{\gamma} u^{\gamma}$.
The null vector can be separated as~\cite{35}
\begin{equation} \label{null vector}
    k^{\alpha} = f \left( u^{\alpha} + n^{\alpha} \right),
\end{equation}
where $n^{\alpha}$ is a vector orthogonal to $u^{\alpha}$.
In the comoving frame, where $u^{\alpha} = (1,0,0,0)$, it follows that $f = - k_{\alpha} u^{\alpha} = dt / d\lambda$.
Using~\eqref{1+3 decomposition},~\eqref{null vector}, and~\eqref{propagation equation}, we obtain the frequency ratio
\begin{equation} \label{frequency ratio}
    \frac{f_{S}}{f_{O}} = \exp \left[ \int_{t_{S}}^{t_O} dt \left( \frac{1}{3} \tilde{\Theta} + \sigma_{\mu\alpha} n^{\mu} n^{\alpha} + \mathcal{I} \right) \right],
\end{equation}
where $\mathcal{I}$ is the torsion-induced correction
\begin{equation} \label{Torsion terms}
    \mathcal{I} = - \frac{2}{3} S_{\nu\mu}{}^{\mu} u^{\nu} + 2 S_{\mu\alpha\nu} n^{\mu} u^{\alpha} \left( u^{\nu} + n^{\nu} \right).
\end{equation}
Here, $f_{S}$ and $f_{O}$ are the frequencies emitted from the source at $t_{S}$ and measured by the observer at $t_{O}$, respectively, so that $f_{S}/f_{O}=1+z$.

For the flat FLRW metric~\eqref{FLRW}, we have $\sigma_{\mu\nu}=0$ and $\tilde{\Theta}=3H$.
Using~\eqref{torsion scalar}, the two torsion contributions in~\eqref{Torsion terms} become
\begin{equation} \label{spin-sourced torsion scalar,,}
    - \frac{2}{3} S_{\nu\mu}{}^\mu u^\nu = \Phi, \qquad
    2S_{\mu\alpha\nu}n^{\mu}u^{\alpha}\left(u^{\nu}+n^{\nu}\right) = -\Phi.
\end{equation}
Thus, $\mathcal{I}=0$.
Solving~\eqref{frequency ratio}, we obtain
\begin{equation} \label{redshift}
    a = \frac{1}{1+z},
\end{equation}
where we have set $a_{S}=a$ and $a_{O}=1$.
Consequently, the standard relation between the scale factor and redshift remains unchanged for the torsion scalar considered here.

\section{Cross-section}
We examine the rate of change of the cross-sectional area of a null bundle in spacetimes with torsion.
Since the naive choice $\hat{h}_{\alpha\beta} = g_{\alpha\beta} + k_\alpha k_\beta$ is not orthogonal to the null vector $k^\alpha$, we introduce another null vector $N^\alpha$. 
In a local Lorentz frame, let $u=t-x$ and $v=t+x$ denote the outgoing and ingoing null coordinates, with $k^\alpha$ and $N^\alpha$ tangent to curves of constant $u$ and $v$, respectively.
The two null vectors satisfy $N^\alpha N_\alpha=0$ and $k^\alpha N_\alpha\neq0$, and may be normalized by $k^\alpha N_\alpha=-1$. 
A projection tensor orthogonal to both null vectors is given by~\cite{39}
\begin{equation} \label{hat h}
    \hat{h}_{\alpha\beta} 
    = g_{\alpha\beta}
    + k_\alpha N_\beta
    + N_\alpha k_\beta.
\end{equation}
It satisfies $\hat{h}_{\alpha\beta} k^\alpha =\hat{h}_{\alpha\beta} N^\alpha = 0$ and $\hat{h}^\alpha_\alpha = 2$, and defines the metric on the two-dimensional screen space.
Using~\eqref{eikonal eq} and~\eqref{null eq}, we obtain the screen-space expansion scalar
\begin{equation} \label{hat Theta 2}
    \hat{\Theta} = g^{\alpha\beta} \hat{h}^\mu_\alpha \hat{h}^\nu_\beta \nabla_\nu k_\mu = \nabla_\mu k^\mu + 2 S_{\mu\nu\rho} k^\mu N^\nu k^\rho.
\end{equation}

To define a cross-section of the null bundle, we select a reference null curve $\gamma(\lambda)$ and a point $P$ on it, where $\lambda=\lambda_P$. 
We also introduce auxiliary curves tangent to $N^\alpha$ and parameterized by $\mu$, with $\mu$ chosen to remain constant along each null curve. 
The auxiliary curve passing through $P$ is denoted by $\beta(\mu)$, and the value of its parameter at $P$ by $\mu_\gamma$.
In a neighborhood of $P$, the two-dimensional cross-section $\delta S(\lambda_P,\mu_\gamma)$ is defined by the intersection of the hypersurfaces $\lambda=\lambda_P$ and $\mu=\mu_\gamma$.
The curves $\gamma$ and $\beta$ are chosen to be orthogonal to the cross-section. 
Since $\mu_\gamma$ is fixed for the chosen reference null curve $\gamma$, we simply denote the cross-section by $\delta S(\lambda_P)$.

We assign coordinates $\theta^A=(\theta^2,\theta^3)$ to each point on $\delta S(\lambda_P)$.
Since exactly one null curve passes through each point, $\theta^A$ label the null curves themselves.
Requiring the coordinates to remain constant along each null curve, points on another cross section $\delta S(\lambda)$ can be identified by the same $\theta^A$.
This construction defines a local coordinate system $(\lambda,\mu,\theta^A)$ near $\gamma$, related to the original coordinates by $x^\alpha=x^\alpha(\lambda,\mu,\theta^A)$.
The null tangent vector and the basis vectors tangent to the cross-section are
\begin{equation}
    k^\alpha 
    = \left( \frac{\partial x^\alpha}{\partial \lambda} \right)_{\mu,\theta^A},
    \qquad
    e^\alpha_A
    = \left( \frac{\partial x^\alpha}{\partial \theta^A} \right)_{\lambda,\mu}.
\end{equation}
Since $\gamma$ and $\beta$ are orthogonal to the cross section, it follows that $e^\alpha_A k_\alpha = 0$ and $e^\alpha_A N_\alpha = 0$.
The Lie derivative of the basis vector along the null curve vanishes, namely
\begin{equation}
    \mathcal{L}_\mathbf{k} \mathbf{e}
    = k^\beta \partial_\beta e^\alpha_A - e^\beta_A \partial_\beta k^\alpha
    = \frac{\partial}{\partial \lambda} \frac{\partial x^\alpha}{\partial \theta^A} 
    - \frac{\partial}{\partial \theta^A} \frac{\partial x^\alpha}{\partial \lambda}
    = 0.
\end{equation}
Therefore, $e^\alpha_A$ serve as connecting vectors.
In terms of the affine covariant derivative, $\mathcal{L}_\mathbf{k} \mathbf{e} = 0$ becomes
\begin{equation} \label{311}
    k^{\beta}\nabla_{\beta}e^\alpha_A 
    = e^\beta_A \nabla_\beta k^\alpha 
    + 2S_{\gamma\beta}{}^{\alpha}e^\beta_A k^{\gamma}.
\end{equation}

On the cross-section $\delta S(\lambda_P)$, an induced metric is defined as $\sigma_{AB} = g_{\alpha\beta} e^\alpha_A e^\beta_B$.
A cross-sectional area is given by $dA = \sqrt{\sigma} d^2 \theta$, where $\sigma = \det
(\sigma_{AB})$ and $d^2 \theta = d\theta^2 d\theta^3$.
The rate of change of the cross-sectional area $dA$ reads
\begin{equation} \label{dA l.h.s}
    \frac{1}{dA} \frac{D(dA)}{d\lambda}
    = \frac{1}{2} \sigma^{AB} \frac{D\sigma_{AB}}{d\lambda}
    = \frac{1}{2} \sigma^{AB} k^\mu \nabla_\mu \sigma_{AB}.
\end{equation}
Using~\eqref{hat Theta 2},~\eqref{311}, and~\eqref{dA l.h.s}, we obtain 
\begin{equation} \label{nabla k}
    \nabla_\alpha k^\alpha
    = \frac{1}{dA} \frac{D(dA)}{d\lambda}
    - 2 S_{\rho\alpha}{}^\alpha k^\rho
    - 4 S_{\mu\nu\rho} k^\mu N^\nu k^\rho,
\end{equation}
where we have used the inverse metric $\hat{h}^{\alpha\beta} = \sigma^{AB} e^\alpha_A e^\beta_B$.


\begin{thebibliography}{99}


\bibitem{1}
A.G. Riess, et al., Observational evidence from supernovae for an accelerating universe and a cosmological constant, Astron. J. 116 (1998) 1009-1038.

\bibitem{2}
S. Perlmutter, et al., Measurements of omega and lambda from 42 high-redshift supernovae, Astrophys. J. 517 (1999) 565-586.

\bibitem{3}
D.N. Spergel, et al., Wilkinson microwave anisotropy probe (WMAP) three year results: implications for cosmology, Astrophys. J. Suppl. 170 (2007) 377.

\bibitem{4}
N. Aghanim, et al., Planck 2018 results. VI. Cosmological parameters, Astron. Astrophys., 641 (2020) A6.

\bibitem{5}
D.J. Eisenstein, et al., Detection of the baryon acoustic peak in the large-scale correlation function of SDSS luminous red galaxies, Astrophys. J. 633 (2005) 560-574.

\bibitem{6}
S. Alam, et al., The clustering of galaxies in the completed SDSS-III baryon oscillation spectroscopic survey: cosmological analysis of the DR12 galaxy sample, Mon. Not. R. Astron. Soc. 470 (3) (2017) 2617-2652.

\bibitem{7}
M. Tegmark, et al., Cosmological constraints from the SDSS luminous red galaxies, Phys. Rev. D 74 (2006) 123507.

\bibitem{8}
T.M.C. Abbott, et al., Dark energy survey year 3 results: cosmological constraints from galaxy clustering and weak lensing, Phys. Rev. D 105 (2) (2022) 023520.

\bibitem{9}
S. Weinberg, The cosmological constant problem, Rev. Mod. Phys. 61 (1989) 1.

\bibitem{10}
S.M. Carroll, The cosmological constant, Rev. Rel. 4 (2001) 1.

\bibitem{11}
P.J. Steinhardt, Critical Problems in Physics, in: V.L. Fitch, D.R. Marlow, Eds. Princeton University Press, Princeton, 1997.

\bibitem{12}
A.G. Riess et al., A 2.4\% determination of the local value of the Hubble constant, Astrophys. J. 826 (1) (2016) 56.

\bibitem{13}
J.-P. Hu, F.-Y. Wang, Hubble tension: the evidence of new physics, Universe 9 (2) (2023) 94.

\bibitem{14}
L. Verde, T. Treu, A.G. Riess, Tensions between the early and the late universe, Nature Astron. 3 (2019) 891.

\bibitem{15}
E. Abdalla, et al., Cosmology intertwined: a review of the particle physics, astrophysics, and cosmology associated with the cosmological tensions and anomalies, JHEAp 34 (2022) 49-211.

\bibitem{16}
A.G. Adame, et al., DESI 2024 VI: Cosmological constraints from the measurements of baryon acoustic oscillations, JCAP 02 (2025) 021.

\bibitem{17}
M. Lucca, Dark energy-dark matter interactions as a solution to the $S_{8}$ tension, Phys. Dark Univ. 34 (2021) 100899.

\bibitem{18}
M. Yashiki, Toward a simultaneous resolution of the $H_{0}$ and $S_{8}$ tensions: early dark energy and an interacting dark sector model, Phys. Rev. D 112 (6) (2025) 063517.

\bibitem{19}
A.G. Cohen, D.B. Kaplan, A.E. Nelson, Effective field theory, black holes, and the cosmological constant, Phys. Rev. Lett. 82 (1999) 4971-4974.

\bibitem{20}
M. Li, A model of holographic dark energy, Phys. Lett. B 603 (2004) 1.

\bibitem{21}
J.W. Lee, H.C. Kim, J. Lee, Holographic dark energy and quantum entanglement, J. Korean Phys. Soc. 74 (1) (2019) 1-11.

\bibitem{22}
S.D.H. Hsu, Entropy bounds and dark energy, Phys. Lett. B 594 (2004) 13-16.

\bibitem{23}
S. Wang, Y. Wang, M. Li, Holographic dark energy, Phys. Rept. 696 (2017) 1-57.

\bibitem{24}
D. Pavon, W. Zimdahl, Holographic dark energy and cosmic coincidence, Phys. Lett. B 628 (2005) 206-210.

\bibitem{25}
E. Cartan, C. R. Acad. Sci. (Paris) 174 (1922) 593. Ann. Sci. Ec Norm. Super. 40, 325 (1923); 41, 1 (1924); 42, 17 (1925).

\bibitem{26}
F.W. Hehl, et al., General relativity with spin and torsion: foundations and prospects, Rev. Mod. Phys. 48 (1976) 393.

\bibitem{27}
F.W. Hehl, P. von der Heyde, G.D. Kerlick, General relativity with spin and torsion and its deviations from Einstein's theory, Phys. Rev. D 10 (1974) 1066.

\bibitem{28}
M. Gasperini, Spin-dominated inflation in the Einstein-Cartan theory, Phys. Rev. Lett. 56 (1986) 2873.

\bibitem{29}
J. Frenkel, Über die von der molekularkinetischen Theorie der Wärme geforderte Bewegung von in ruhenden Flüssigkeiten suspendierten Teilchen, Z. Physik. 35 (1926) 652-669.

\bibitem{30}
S.B. Medina, M. Nowakowski, D. Batic, Einstein-Cartan cosmologies, Ann. Phys. 400 (2019) 64-108.

\bibitem{31}
M. Tsamparlis, Cosmological principle and torsion, Phys. Lett. A 75 (1979) 27-28.

\bibitem{32}
M. Tsamparlis, Methods for deriving solutions in generalized theories of gravitation: the Einstein--Cartan theory, Phys. Rev. D 24 (1981) 1451-1463.

\bibitem{33}
D. Camarena, V. Marra, Local determination of the Hubble constant and the deceleration parameter, Phys. Rev. Res. 2 (2020) 1, 013028.

\bibitem{34}
J.W. Lee, J. Lee, H.C. Kim, Dark energy from vacuum entanglement, JCAP 08 (2007) 005.

\bibitem{35}
K. Bolejko, M. Cinus, B.F. Roukema, Cosmological signatures of torsion and how to distinguish torsion from the dark sector, Phys. Rev. D 101 (10) (2020) 104046.

\bibitem{36}
G.F.R. Ellis, Republication of: relativistic cosmology, Gen. Rel. Grav. 41 (2009) 581-660.

\bibitem{37}
R. Fresneda, M.C. Baldiotti, T.S. Pereira, Maxwell field with torsion, Braz. J. Phys., 45, (3) (2015) 353-358.

\bibitem{38}
I.L. Shapiro, Physical aspects of the space-time torsion, Phys. Rept., 357, (2002) 113.

\bibitem{39}
E. Poisson, A Relativist's Toolkit: The Mathematics of Black-Hole Mechanics, Cambridge University Press, Cambridge, 2004.

\end{thebibliography}
\end{document}